# Narrow-Band Pulsed Electron Source Based on Near-Threshold Photoionization of Cs in a Magneto-Optical Trap

O. Fedchenko[1], S. Chernov[1], G. Schönhense[1], R. Hahn[2], and D. Comparat[2]

1 Institut für Physik, Johannes Gutenberg-Universität, 55128 Mainz, Germany

2 Laboratoire Aimé Cotton, CNRS, Université Paris-Sud, ENS Paris-Saclay, Université Paris-Saclay, Bat.505, 91405 Orsay, France

The newly developed method of time-of-flight (ToF) momentum microscopy was used to analyse the cold electron emission from a Cs 3D magneto-optical trap (MOT). Three-step resonant photoionization was implemented via two intermediate states ($6P_{3/2}$ pumped with 852 nm laser and $7S_{1/2}$ with 1470 nm) and a tuneable femtosecond Ti:sapphire laser for the final ionization step. The magnetic field of the MOT is switched off during the photoionization step. The natural bandwidth of the fs-laser is reduced to 4 meV using optical spectral filters. Precise tuning of the photon energy makes it possible to observe the transition regime between direct photoemission into the open continuum and field induced ionization of highly-excited Rydberg states. The paths can be identified by their characteristic dependency on the extraction field and on the Ti:sapphire polarization. ToF analysis allowed us to disentangle the ionization paths and the dependence of the spatio-temporal distribution of the cold electrons on the polarization of the ionizing laser.

## I. INTRODUCTION

Ultracold electron sources could be eminently suitable for the investigation of ultrafast processes with electron diffraction due to highly coherent electron bunches from the near-threshold photoionization of laser-cooled atoms [1,2-4]. This type of sources could potentially yield an increase in brightness by reducing the temperature of the electrons [1]. Furthermore, such sources can employ large active areas (hundreds of micrometres) that allows extracting a large number of electrons to record the diffraction patterns [5, 6]. Typically ~$10^8$ atoms could be trapped in such a cold cloud with rms radius of ~1 mm, at a source temperature of $T$ ~ 150 μK [7,8]. The loading time of the magneto-optical trap is around 100 ms and depends on the experimental conditions [9]. The lifetime of the $6^2P_{3/2}$ state that is pumped with the MOT-lasers is 30 ns [10].

There are several ways of producing very monochromatic electron beams, for example, using anion photo-detachment spectroscopy with combination of slow electron velocity-map imaging [11,12]. This has the advantage of having no Coulomb interaction between the ejected electron and the atomic core which is neutral, but the Coulomb interactions between the negative ions in the beam poses a serious limit to the achievable photoelectron flux.

The possibility of formation of high-coherence electron bunches from an ultracold electron source even with large bandwidth of femtosecond photoionization pulses was demonstrated in [9]. Since the temperature is one of the key parameters to characterise an electron source based on a magneto-optical trap (MOT), all factors that have effect on it should be investigated [4]. The polarization of the ionization laser is among them. The influence of polarization of ionization laser on the effective temperature of electrons, emitted from a MOT was studied for the case of near-threshold photoionization [13]. All factors, influencing the temperature, also affect the differential ionization cross-section and spatio-temporal distributions of the photoelectrons [14]. As discussed in [15] it is very important to control the polarization to form electron bunches with the highest transversal coherence length that determines the condition of the ultrafast electron sources for high-quality diffraction. Alongside with the large coherence demanded for diffraction, a small energy bandwidth is required for high-resolution electron energy-loss spectroscopy and short pulse lengths are attractive for the study of dynamic processes.

In the present experiment, we add to the previous studies the investigation of the effect of the polarization below, near and above (zero field) threshold excitation using femtosecond-laser excitation of a cold Cs MOT with magnetic field turned off. The spatio-temporal and spectroscopic analysis of the electron beam is performed with a high-time-resolution detector (~200 ps) and the Time-of-Flight (ToF) momentum microscopy technique. This is the follow-up of our previous work [8] that was conducted in the presence of a static magnetic field. Under those conditions the magnetic field broadened the ToF and induced a complicated mode structure due to spiral trajectories [8]. In the current paper an improved experimental setup with a rapidly switched magnetic field is presented. Furthermore, the effect of the laser polarization on the excitation process above and below zero-field threshold using fs-laser pulses with broad (24 meV) and narrow bandwidth (4 meV) is described. The influence of the orientation of the electric vector relative to the electrostatic extraction field on the spatio-temporal distribution of the electrons is substantial for above-threshold excitation but less significant for excitation below threshold.

## II. EXPERIMENTAL SETUP

The experimental setup is similar to the one used in our previous work [8] but now with the magnetic field switched off during the pulse of the ionizing laser, see Fig. 1. This allowed us to investigate the influence of the polarization of the ionizing laser on the spatio-temporal distribution of photoelectrons from the Cs atom cloud. Briefly, to extract the electrons from the Cs MOT, a three-step resonant photoionization process was implemented via two intermediate states, $6P_{3/2}$ pumped with a 852 nm laser and then $7S_{1/2}$ via a 1470 nm laser. The ionization step was implemented by a tuneable Ti:sapphire laser allowing direct photoionization above threshold or the resonant excitation of high-lying Rydberg states a few meV below threshold (with subsequent field ionization). The intermediate states were populated by a CW beam, the ionizing laser provides fs pulses at a repetition rate of 80 MHz. The femtosecond laser pulses are usually Fourier-transform limited and in our case an optical spectrometer revealed an intrinsic bandwidth of 24 meV FWHM (at 835 nm) and 0.8 W average laser power.

At first, electrons are accelerated by a low constant field of ~2 V/cm and then enter a 1 m long drift tube that converts spatial and energy spread into time spread. At the end of the drift tube the electron signal is recorded by a Delay-Line Detector (DLD) [16] that contains 2 MicroChannel Plates (MCP) as pulse intensifier. As an extension to our previous experiment an electrical circuit for switching on/off the 1470 nm excitation laser and magnetic coils of the MOT was added. The magnetic-field switching is conducted at a frequency of 100 Hz, with coils' current turned off every 10 ms for a duration of 5 ms. After the coils were off for 4 ms the acousto-optical modulator (AOM) for the excitation laser (1470 nm) is switched on for 1 ms and the signal from the DLD

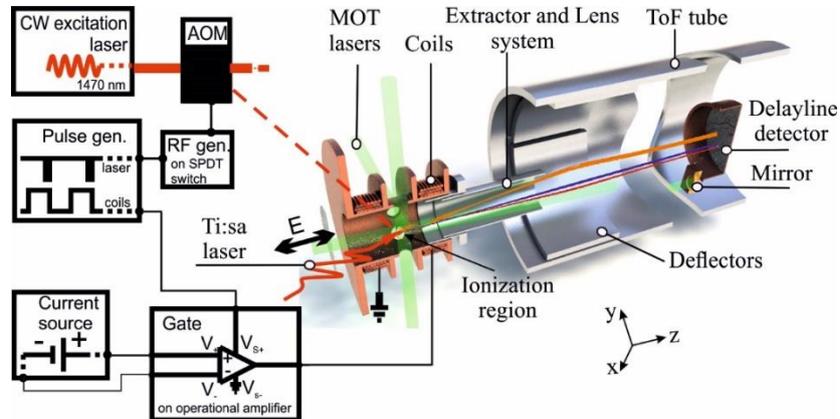

FIG. 1. Schematic view of the electron-optical extraction and imaging system (right part) and electrical circuit for switching the AOM for the 1470 nm excitation laser and the magnetic field (left part).

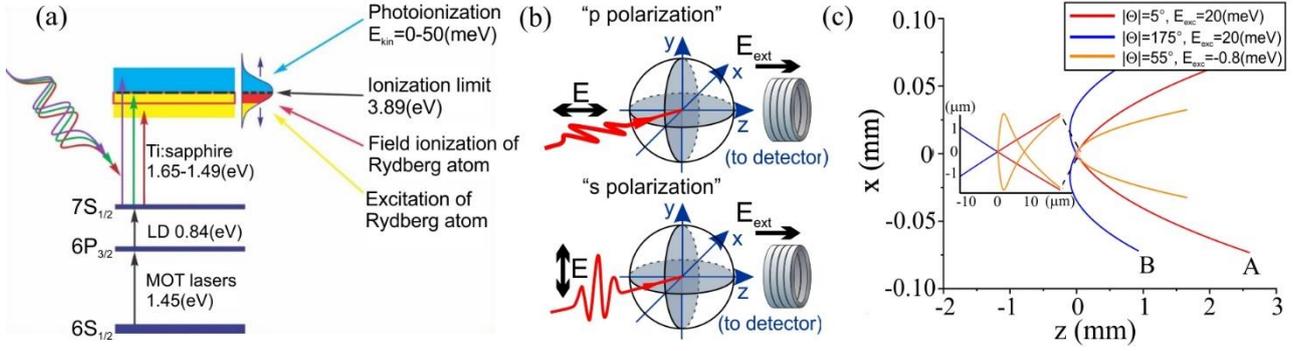

FIG. 2. (a) Scheme of ionization paths that can take place simultaneously. (b) Experimental geometry including the photon polarization; $E$ denotes the electric vector of the ionization laser, $E_{ext}$ the static extractor field of the electron optics. (c) Three examples of classical trajectories, simulated over 10 ns, of electrons escaping the potential of the Cs$^+$ ion under an electric field of 2 V/cm, with various initial ejection angles θ and various kinetic energies $E_{exc}$. See text for details of the simulation. The inset displays the magnified area near the emitter atom.

is being recorded.

The detection system (DLD) was synchronised with the arriving photon pulses by using the 80 MHz trigger signal of the Ti:sapphire laser. To avoid detection of background electrons the excitation laser was switched off with a repetition rate of 100 Hz. As a result, only one single laser pulse was selected.

The delay of 4 ms before the measurement effectively eliminates the influence of residual magnetic fields caused by eddy currents in the anti-Helmholtz coils. The tests show that during the first 3 ms the image on the DLD appears to be distorted, that means that photoelectrons still "feel" the influence of the Lorentz force.

The same effect was observed in our previous experiments, where the spatio-temporal distribution of electrons revealed spiralling trajectories [8].

The measurements with different polarization of the ionization laser were done with the help of a half-wave plate that is able to produce either p-polarization (electric-field vector parallel to the acceleration axis) or *s*-polarization (electric-field vector perpendicular to the acceleration axis), see Fig. 2(b). The results also depend on the electron energy that is controlled with the Ti:sapphire laser bandwidth and central frequency. The presence of a low electric field (around 2 V/cm) decreases the classical ionization threshold (3.895 eV in zero-field) to 3.894 eV (1.596 and 1.595 for the Ti-sapphire laser).

Fig. 2(a) presents the three types of excitation and electron ejection that may occur:
- Photoionization via excitation above zero-field threshold (in blue).
- Field-ionization of Rydberg states (in red). This correspond to an excitation of Rydberg states that would have been stable under the absence of field but are ionized due to the lowering of the ionization barrier because of the presence of the external field.
- Excitation of Ryberg states below the classical field-ionization threshold (in yellow). In this region we excite Rydberg states that are in principle stable [17] but could be ionized for example by absorbing a 300 K black-body photon (thermal radiation).

### III. RESULTS

The first measurements were performed for various nominal photon energies of the ionizing laser in the range of -19 meV to +19 meV from the zero-field photoemission threshold. Time-of-flight spectra are presented in Fig. 3.

From the results it is clearly seen that ToF electron distributions are significantly different for *p*- and *s*-polarizations.

This is particularly visible for above-threshold ionization, see Fig. 3(c,f). In this region electrons have essentially classical trajectories [18] and it is quite easy to understand the temporal distribution of the photoelectrons in this classical picture. For instance, for p-polarization the laser field oriented parallel to the field axis (see Fig. 2(b)) produces two bunches of electrons: first− the electrons emitted at small emission angles (around $\theta = 0°$) to the optical axis and second− the ones emitted backwards (close to $\theta = 180°$). The electrons emitted away from the DLD spend more time (because of their longer path) in the MOT region before they are turned around by the extraction electric field [8, 19], that is why they arrive later, forming the peak labelled with *B* in Fig. 3(c), compared to the one ejected directly toward the detector, forming the peak labelled with *A* in Fig. 3(c). The simulations to confirm these explanations were performed in terms of classical trajectories. The General Particle Tracer [20] and the formalism introduced in Ref. [21] were used for the first 10 ns for the electrons escaping the modified 1/r potential of $Cs^+$ in a 2 V/cm electric field. The results for the same high initial kinetic energy (corresponding to the experimental central wavelength in Fig. 3(c)) but with two different absolute values of the starting angle ($|\theta| = 5°, 175°$) are presented in Fig. 2(c) in red and blue, respectively. The trajectory for the electron with low energy (just sufficient for ionization) with a starting angle of $|\theta| = 55°$ is shown in Fig. 2(c) in orange colour.

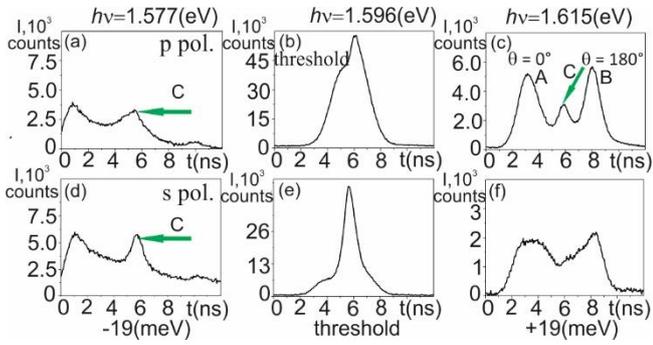

FIG. 3. Time-of-flight spectra taken at 3 photon energies close to the photoionization threshold with a wide bandwidth of the Ti:sapphire laser (ΔE= 24 meV). (a-c) with p-polarized light; (d-f) same but for *s*-polarized light.

These simulations give us a good insight on the composition of the three experimental peaks (labelled *A*, *B*, and *C* in Figs. 3 and 2). Indeed, the experimental delay found between those peaks is explicable by the different positions of electrons *A*, *B* and *C* after 10 ns.

The delay of arrival time ($\tau$) between the two maxima (peak *A* and *B*) are directly related to the extracting electric field $E_{ext}$ and to the initial velocity $v_0$ of the emitted electron. The gap between the two peaks becomes less pronounced when the change of the laser frequency gives less excess energy for the emitted electrons. The formula $\tau = \frac{2v_0 m}{eE}$ (where *m* and *e* are the mass and charge of the emitted electron, respectively) can be used to estimate an extraction field of 2 V/cm in the ionization zone. The field value is indeed in a good agreement with the one we get with Simion simulations of our geometry (despite not precisely known the position of the extractor and lens system shown in Fig. (1)) [22].

The experimental signals (such as peaks *A* and *B*) are broadened due to the difference in initial kinetic energies (due to the 24 meV laser bandwidth), but also because of several ejection angles $\theta$ (with respect to the polarization vector of the incident light). This is expressed by the standard formula of the photoionization differential cross-section $d\,d\sigma/d\Omega = (\sigma/4\pi)[1+ \beta (3 \cos^2(\theta)-1)/2]$ (with β being the asymmetry parameter [12]). As we start from the 7s state the free electron is emitted in a *p*-partial wave, so β β = 2 [23, 24]. This leads to a "∞-shape" of the photoionization cross-section for *p*-polarization and to "8-shape" in case of *s*-polarization, explaining the shape of the observed peaks.

In addition to signals *A*, *B* that clearly vary with photon energy as expected, there is another sharp and intense line in the time histograms that shows a fundamentally different behaviour (see peak *C* marked with the green arrows in Fig. (3)). The signal has the highest intensity close to the ionization threshold and remains present even with the decrease of laser energy down to 18 meV below the ionization limit (Figs. 3(a,d)). This signal can be attributed to the excitation of rapidly ionizing Rydberg states below the zero-field ionisation threshold (red region in Fig. 2(a)). Thanks to the broad bandwidth of the laser, these states are addressed even when the laser is centred far away from the threshold. The broadening of peak *C* does not seem to be affected by the laser polarization (only its intensity is). A simple classical picture (cf. Fig. 2(c) and the inset of this figure) shows that even if the Rydberg electron trajectory initially depends on the laser polarization, the combined influence of the ion core potential and electric field on the very low energy electron is sufficient to deeply modify the trajectory. For instance, as studied in photoionization microscopy the glory effect tends to produce an emission mostly in the direction of the field even if the initial emitted angle is not in this axis [14,17,18].

Finally, the electron signal resulting from the excitation of the lower-lying Rydberg states, corresponding to the yellow region in Fig. 2(a), usually lies in the μs time range (due to long tunneling processes or black-body radiation-induced ionization) [1,25,26], also produces the high background visible on the experimental results.

In order to confirm these hypotheses, the extraction field was varied between 1 and 4 V/cm. Results are presented in Fig. 4. The main effect is the increase of the intensity of peak C during field increase. Indeed because of the wide laser bandwidth many Rydberg states are excited and with the increase of extraction field more of them are ionized, leading to an increase of the total intensity.

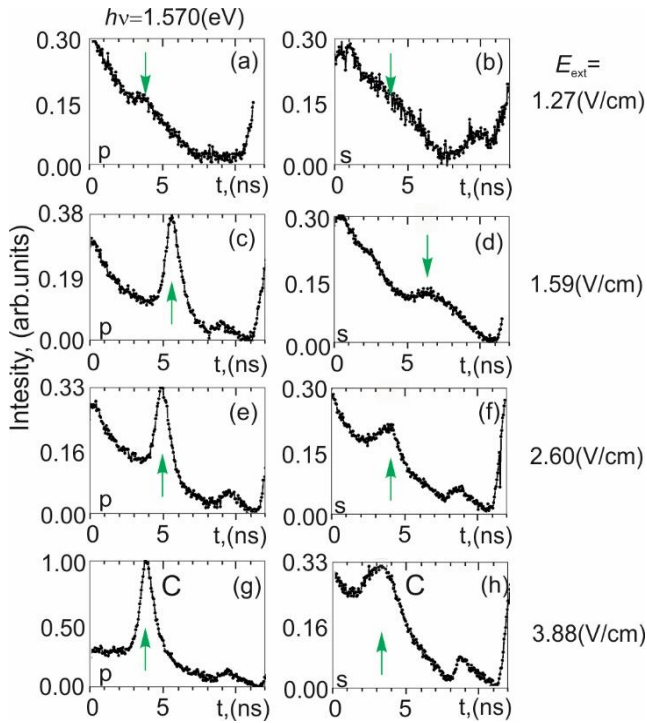

FIG. 4. Dependence of the field ionization spectra on the strength of the extraction field. Measurements were performed with ionization-laser bandwidth $\Delta E$= 24 meV, energy below threshold -19 meV. (a,c,e,g) p-polarized light, (b,d,f,h) s-polarized light.

Due to the large laser bandwidth ($\Delta E = 24$ meV FWHM) the total signal (see for instance Fig. 3(c)) is the superposition of signals originating from different ionization paths, photoionization with peaks A and B and field ionization of high-lying Rydberg states with peak C. In order to clarify the picture we took a second set of measurements with much narrower bandwidth of the laser. This was achieved with the help of variable long- and short-pass filters for the Ti:sapphire laser. These filters allow to reduce the bandwidth to 2 nm which corresponds to $\Delta E$= 4 meV. The spatio-temporal electron distributions obtained in this case are shown in Fig. 5.

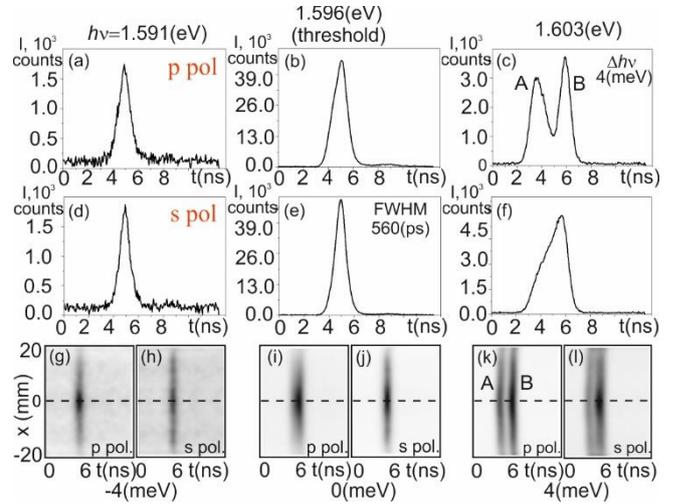

FIG. 5. (a-c) Time-of-flight spectra taken at 3 photon energies close to the photoionization threshold with p-polarized light in case of narrow bandwidth of the Ti:sapphire laser ($\Delta E$= 4 meV). (d-f) Same but for s-polarized light. Extractor field $E_{ext}$ = 2 V/cm, B-field switched off during ionization step. (g,i,k) Spatio-temporal distribution of the electron emission signal on the delay-line detector in 2D space-time sections through the optical axis (dashed line) for the measurements with p-polarization. (h,j,l) Same for s-polarization.

The reduction of the laser bandwidth to 4 meV strongly confines the temporal distributions. The situation is now more clear, for instance, with no Rydberg excitation at 4 meV above threshold (see Figs. 5(c,f)), because for the bandwidth of $\Delta E$= 4 meV Rydberg states are not populated and peak C is not present.

At threshold excitation (Figs. 5(b,e,i,j)) peaks A and B overlap and s- and p-polarizations yield similar distributions. With s-polarization (Fig. 5(e)) we found a width of 560 ps FWHM only, which is less than factor of 3 from the time-resolution of the delay-line detector (200 ps). This is a good initial situation in order to produce an ultra-cold electron beam.

The difference between p- and s-polarization is now clearly visible above threshold (compare Figs. 5(c) and 5(f)). In case of s-polarization the "8-shape" of the photoionization cross-section causes ejection of electrons preferably perpendicular to the optical axis, but the two opposite emission directions are (for the time of flight) equivalent and only one electron bunch is observed at the DLD. This again, in contrast to peaks *A* and *B* for the p-polarization, corresponds to forward and backward emission with respect to the extractor-field direction. Another indication of the electron trajectories is given by the electron position recorded on the detector. Namely, the higher value of the electrons' initial energy, the larger diameter of the final spot (compare for example Figs. 5(i) and 5(k)). Furthermore, the final spot is sometimes clipped by 60 mm diameter of the DLD (see Fig. 5(l)) indicating that some electrons simply do not reach the final detector under the very low electrostatic extraction field applied. This precludes a detailed study of the relative peaks intensities.

The data shown in Fig. 5 are still somewhat broadened because of the initial spatial spread (due to the finite laser spot). This spatial spread leads to an initial energy spread due to the differential potential across the ionization zone.

## IV. SUMMARY AND CONCLUSION

In order to obtain a pulsed ultracold electron source with few-meV bandwidth we used three-step resonant laser photoionization of Cs atoms in a magneto-optical trap (MOT). A tuneable Ti:sapphire laser with bandwidth reduced by optical spectral filters allows to selectively access three different emission regimes: photoionization with excess energy of few meV, at-threshold ionization and the Rydberg regime for which even weak electric fields of ~ 2 V/cm lead to effective field ionization. The absence of the Lorentz force without magnetic field leads to much simpler trajectories (without spiral movement) and hence to much narrower time and energy spread than was observed in our earlier work with static field [8]. The measured spectra show a pronounced asymmetry for different polarizations of the ionization laser. This effect, clearly identified with simulations, was not seen in other cold-electron source experiments [1,25] because the initial energy spread due to the high extracting fields blurred the temporal signals. The very low extracting field of the present setup allowed to observe such temporal structures. The significant difference between s- and p-polarization shows that the spatial distribution in Rydberg state plays a role for the field-ionization probability. Reduction of the photon bandpass to 4 meV refines the distributions towards narrower energy and time bandwidths. In the optimal case (at-threshold ionization with s-polarized light) the temporal width was found to be 560 ps FWHM.

In the future even narrower electron bunches could be generated from cold atoms using pulsed electric field ionization of highly-excited atoms. Pulsing the field, for instance by turning it off before the electrons leave the acceleration area, can also be useful for many other applications such as aberration corrections, beam bunching, modification of the longitudinal energy spread, focusing of the beam and more [27-31].

## ACKNOWLEDGMENTS

The present project received funding from the European Research Council under the grant agreement n. 277762 COLDNANO and ANR/DFG (SCHO 341/14-1) HREELM. We would like to express our gratitude to F. Schmidt-Kaler (Univ. of Mainz) for the help with the MOT implementation.